# 5G and Beyond: Smart Devices as part of the Network Fabric

B. Coll-Perales, J. Gozalvez and J.L. Maestre
UWICORE, Ubiquitous Wireless Communications Research Laboratory, http://www.uwicore.umh.es
Universidad Miguel Hernandez (UMH) de Elche, Avda. de la Universidad, s/n, 03202 Elche, Spain
bcoll@umh.es, j.gozalvez@umh.es

**Abstract–** 5G networks mainly rely on infrastructure-centric cellular solutions to address data traffic and service demands. Continuously scaling infrastructure-centric cellular networks is not exempt of challenges, and beyond 5G networks should consider the active coexistence and coordination of infrastructure-centric and device-centric wireless networks. Device-centric wireless networks will build from device-to-device communications (D2D) and multi-hop cellular networks (MCNs). Device-centric wireless networks can push the limits of edge computing and networking to smart devices (including smartphones, vehicles, machines and robots), and exploit their mobile computing, storage and connectivity capabilities. These capabilities can be more efficiently utilized using demand-driven opportunistic networking that establishes the connections between devices and nodes not just based on their presence, but also on their capacity to support the requested demand and services. This paper presents results from experimental field tests that demonstrate the cellular spectral efficiency gains that can be achieved from the combined use of device-centric wireless communications and demand-driven opportunistic networking. The field trials demonstrate that these technologies can improve the cellular spectral efficiency of conventional cellular communications by up to a factor of 4.7 and 12 in outdoor pedestrian and vehicular scenarios, respectively, under the evaluated scenarios and conditions. These gains have been obtained using 4G and IEEE 802.11 technologies. However, the potential of device-centric wireless networks is not constrained to any particular radio interface. The results presented in this paper provide empirical evidences that further motivate progressing towards a new paradigm where edge networking capabilities are moved to smart mobile devices that become part of the network fabric, and can opportunistically and locally integrate network management functions to ensure that sufficient resources are placed where the demand arises.

## 1. Introduction

5G mainly relies on infrastructure-centric solutions that reduce the infrastructure cost per bit and facilitate ultra-dense networks with network nodes deployed as close as possible to the end devices. Addressing future data traffic demands by only scaling current infrastructure-centric cellular solutions is not exempt of challenges, and beyond 5G networks should consider complementing infrastructure-centric solutions with edge-centric computing and networking ones [1]. A further evolution is device-centric networks [2] that move the edge of the network to the smart devices, and exploit their mobile computing, storage and connectivity capabilities. These devices include smartphones, as well as connected vehicles, machines or robots that do not have the energy constraints of smartphones. In device-centric scenarios, smart mobile devices will not only consume or generate data, but will also actively participate in the network management and operation through a carefully designed cooperation and coordination with the cellular infrastructure. Device-centric wireless solutions include device-to-device (D2D) communications and multi-hop cellular networks (MCNs). MCNs are also referred to as UE (User Equipment)-to-network relay in 3GPP standards. MCNs integrate D2D and cellular communications, and allow mobile devices to connect to the cellular infrastructure through other intermediate mobile nodes. MCN is part of the 3GPP roadmap, e.g. to connect and manage machine-type communication devices [3]. D2D is also recognized as a core component of 5G by the Expert Advisory Group of the European Technology platform Networld 2020 [4]. This Group also identifies in its Strategic Research and Innovation Agenda the potential of MCN in beyond 5G where end-user terminals will act as elements of the network to provide connectivity to other terminals [4]. These terminals could actually include vehicles that act as moving cells or moving relay nodes to connect passengers to the Donor eNB as proposed by 3GPP in [5]. Previous experimental studies have shown the benefits that MCNs using D2D communications can provide over conventional cellular systems in terms of QoS, capacity and energy consumption [6]. For example, [6] showed that MCN can reduce by 50% the total energy consumption when comparing a 2-hop uplink MCN connection to a conventional cellular link in an indoor to outdoor scenario. An important aspect for the success of D2D and MCN is securing their communications [7]. To this aim, 3GPP maintains a technical specification on security aspects for Proximity-based Services that addresses both the D2D and MCN communication modes [8].

Opportunistic networking can also play a relevant role in future networks [4]. Opportunistic networking was initially proposed for disconnected networks [9]: in the absence of forwarding opportunities, mobile devices would store the message and carry it until new forwarding opportunities appear. Relevant studies ([10] and [11]) have proposed to extend the concept of opportunistic networking to situations where connections between devices and nodes (other devices or the BS) are established not just based on their presence, but also on their capacity to support the requested service or network demand. We refer to this extension as demand-driven opportunistic networking. The demand can be, for example, to guarantee a given latency, reliability or throughput level, to minimize the energy consumption, or to maximize the spectrum efficiency and the network capacity. The concept of demand-driven opportunistic networking can hence be tailored to different user, service or network demands. In this paper, the demand is the cellular spectral efficiency, and the demand-driven opportunistic networking protocols are configured to increase the cellular spectral efficiency. To this aim, we exploit the fact that mobile data traffic will be dominated by non-real-time traffic (e.g. video streaming that buffers data equivalent to a few tens of seconds of playback), and hence we utilize the time available until the service deadline to search for connections that can improve the transmission efficiency with no cost to the end-user quality of experience.

Demand-driven opportunistic networking offers the possibility to cooperatively exploit the resources of smart mobile devices (with the assistance and coordination of the network) to satisfy demands. This is particularly the case under urban environments where each device has multiple connection options simultaneously. [10] and [11] proposed to combine opportunistic networking and device-centric wireless communications (in particular MCN), and demonstrated analytically and through simulations how such combination can significantly improve the energy-efficiency and capacity of cellular networks. In particular, [10] theoretically analyzed the maximum capacity gains that opportunistic MCN networks can achieve over conventional cellular communications. The study shows that opportunistic MCN can benefit from the mobility of relays to perform transmissions close to the BS where higher data rates can be utilized. Higher data rates reduce the usage of cellular radio resources, and hence can augment the capacity. In [11], the authors analytically demonstrate that the integration of opportunistic networking and MCN communications can reduce the energy consumption by more than 90% compared to conventional cellular communications. In [12], the authors derive the upper-bound capacity gain that can be achieved with 2-hop uplink opportunistic MCN communications compared to conventional cellular communications. The study shows that the capacity gains augment with the service transmission deadline and the cell radius. For example, the study shows that opportunistic MCN can increase the capacity compared to conventional cellular communications by a factor of up to 4.5 with a cell radius of 770m. These results highlight the potential of combining demand-driven opportunistic networking and device-centric wireless communications. The potential for such combination has also been acknowledged by the Expert Advisory Group of the European Technology platform Networld 2020 that provides the following recommendation for beyond 5G networks [4]: "*The combination of novel communication paradigms, including multi-hop D2D, delay tolerant networking where humans, animals and vehicles will be used as data carriers, should be fostered, in order to achieve the maximum degree of flexibility for the network, making it able to adapt to the ever-changing spatial-temporal characteristics of traffic demand*".

This paper progresses the current state of the art by presenting, to the authors' knowledge, the first field trials that experimentally demonstrate the cellular spectral efficiency gains that can be obtained from exploiting device-centric wireless communications and demand-driven opportunistic networking. The trials have been conducted in real pedestrian (outdoor and indoor-to-outdoor) and vehicular scenarios in the city of Elche in Spain. The obtained results provide empirical evidences that further motivate progressing towards a Beyond 5G vision where mobile devices can become moving edge nodes of the network and be part of the network fabric.

## 2. Testbed

A testbed has been implemented to analyze the benefits that device-centric wireless communications and demand-driven opportunistic networking can provide to future cellular networks. The testbed focuses on uplink communications to have close control of the transmission processes. It is used to compare the performance obtained under the following communication modes (Fig. 1.a): 1) conventional cellular (CC) connections where the source node (SN) directly transmits the data to the base station (BS), 2) opportunistic CC connections where the SN exploits the demand-driven opportunistic networking principle to decide when to upload the information to the BS (in $t_1$ in Fig. 1.a), 3) MCN connections that integrate D2D and cellular communications, and use mobile devices as relays to transmit the data to the BS, and 4) opportunistic MCN connections where both the source and relay nodes exploit the demand-driven opportunistic networking principle to decide when to transmit packets (in Fig. 1.a, SN transmits the data to RN in $t_0$, while RN waits until $t_1$ to upload the data to the BS).

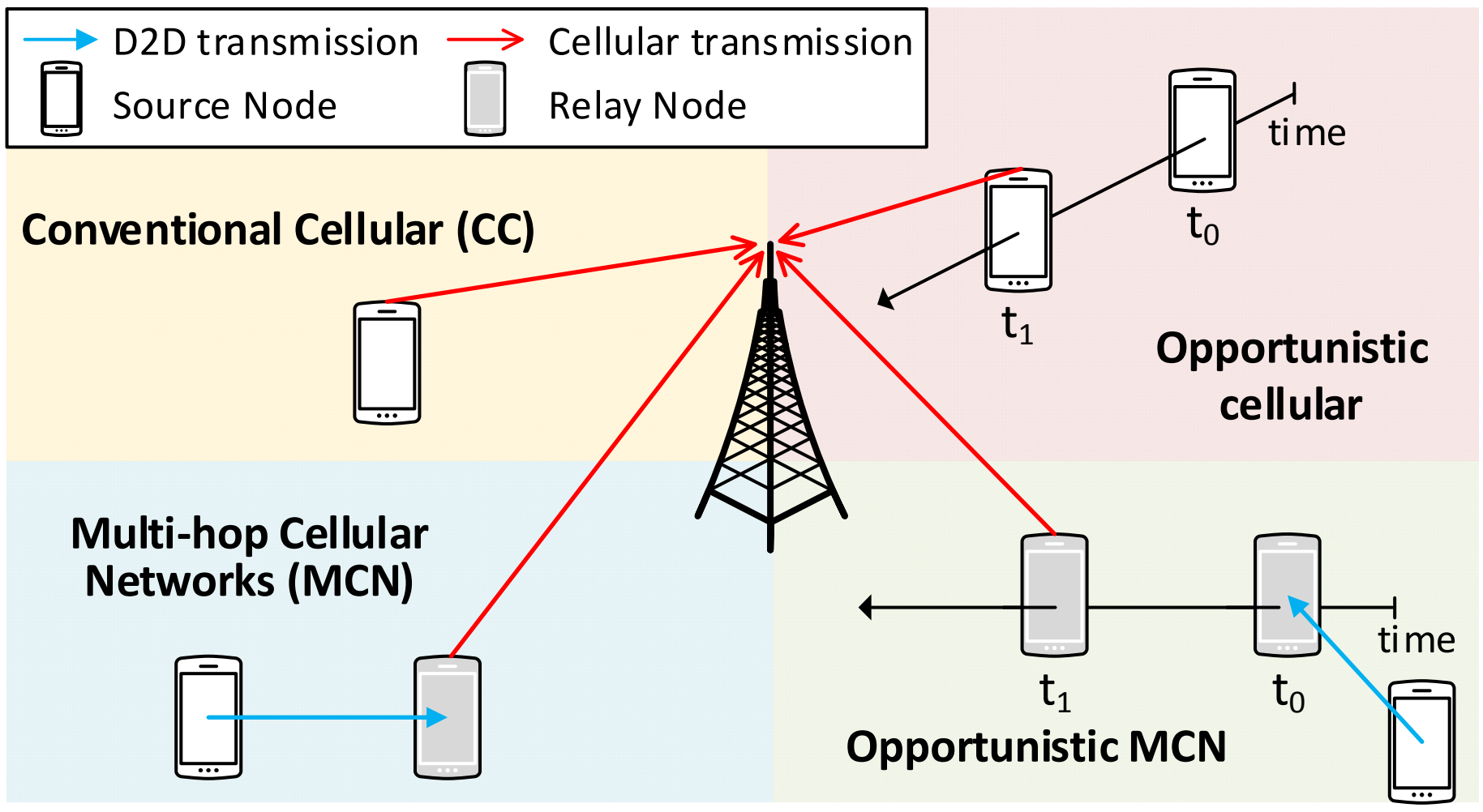

a) Communication modes under evaluation: Conventional Cellular (CC), Multi-hop Cellular Networks (MCN), Opportunistic cellular (Opp CC) and Opportunistic MCN (Opp MCN).

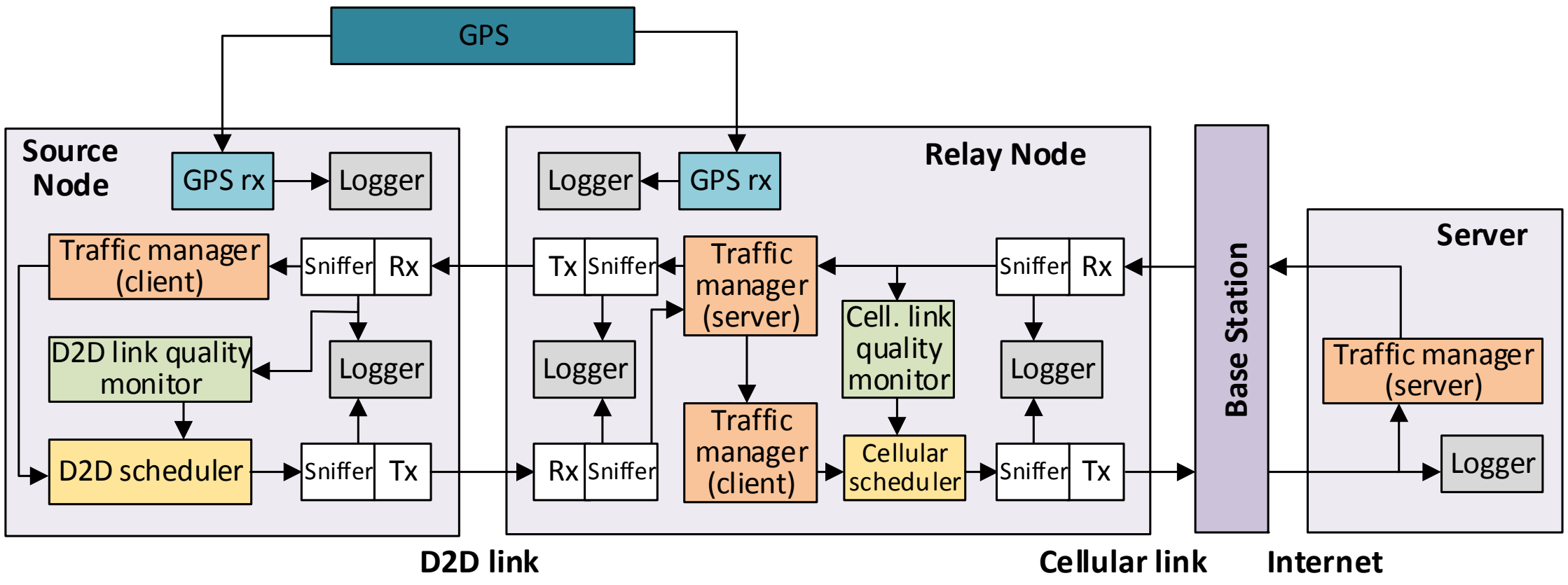

b) Implemented testbed for the opportunistic MCN communication mode

**Fig. 1.** Communication modes (a) and implemented testbed (b).

Fig. 1.b shows the architecture and modules of the testbed when configured to establish a 2-hop opportunistic MCN uplink connection. The mobile nodes in the testbed are laptops. The SN communicates with the relay node (RN) using 802.11g at 2.4GHz or 802.11a at 5GHz[1]. The cellular link from the RN to the BS is established using 4G LTE (Long Term Evolution) at 1.8GHz. For this link, the RN uses a LTE Cat. 3 smartphone. All mobile devices include a GPS receiver ('GPS rx' in Fig. 1.b) used to track their position and time-/geo-reference all the measurements. The nodes capture the exchanged packets using packet sniffer tools ('Sniffer' in Fig. 1.b) developed by the authors. The packets captured are logged for post-processing ('Logger' in Fig. 1.b), and are utilized in real-time to monitor the quality of the D2D and cellular links ('D2D/Cellular link quality monitor' in Fig. 1.b).

In the current implementation, the SN wants to upload a file to a server located at the University. The testbed implements a client-server traffic service for each hop. The client generates the packets to be transmitted, and the server periodically sends reports to its client to indicate which packets were correctly received. At the RN, the server of the D2D link with the SN is connected to the client of the cellular link so that the cellular link cannot transmit more packets than received by the RN through the D2D link. The D2D and cellular transmissions are scheduled as a function of their link quality that is provided by the 'D2D/Cellular Link quality monitor' modules. The current design allows for the independent scheduling of the D2D and cellular transmissions at the SN and the RN.

[1] 3GPP considers both IEEE 802.11 and LTE for D2D (or *sidelink* as referred to in 3GPP) communications. 2.4GHz is used for the outdoor trials and 5GHz for the indoor-to-outdoor trials.

The 'D2D link quality monitor' module in SN monitors the received signal strength indicator (RSSI) of the D2D link using the received control and management packets from the RN, in particular the ACK and beacons respectively. Following [13], D2D transmissions are scheduled based on the RSSI of the D2D link. The 'D2D link quality monitor' module computes the average RSSI (*RSSIavg*) over the last *Nb_Rx* received packets, and passes this information to the 'D2D scheduler'. The scheduler pauses the D2D transmissions if a number *Nb_BelowThr* of consecutive *RSSIavg* values are below the threshold *RSSIthr*. If a D2D link is paused, its quality is still monitored at the SN using the management packets received from the RN. These packets are transmitted even if no data packets are exchanged over an established IEEE 802.11 D2D link. Using the information received from the 'D2D link quality monitor' module, the 'D2D scheduler' module checks every *T_D2D* seconds whether the D2D link quality conditions have improved. When *RSSIavg* is equal or higher than *RSSIthr*, the 'D2D scheduler' resumes the exchange of data packets over the D2D link.

Cellular transmissions from the RN to the BS are also opportunistically scheduled based on their link quality and efficiency. To this aim, a sniffer has been developed to capture the packets exchanged over the cellular link. The sniffer reads from the diagnostic port of the cellular chipsets. Some of the LTE quality metrics collected in real-time by the sniffer include: reference signal received power (RSRP), modulation, number of physical resource blocks (PRBs), and transport block size index ($I_{TBS}$). The RN uses the $I_{TBS}$ and RSRP metrics to opportunistically schedule its uplink cellular transmissions. The $I_{TBS}$ varies between 0 and 26 [14] with the highest values associated to the highest transmission modes (i.e. the combinations of modulation and coding schemes resulting in the highest data rates) and larger transport block sizes. The BS notifies the mobile station about which resources (PRBs) and $I_{TBS}$ it should use for its uplink cellular transmissions. This notification is included in the downlink control information (DCI) reports when there is an active uplink cellular connection. The 'cellular link quality monitor' module continuously estimates the average $I_{TBS}$ ($I_{TBS}_avg$) over the last *TcellAvg* seconds. If $I_{TBS}_avg$ is below a threshold $I_{TBS}_thr$, the 'cellular scheduler' module pauses the uplink cellular data transmissions. This opportunistic management of the cellular transmissions benefits the cellular spectral efficiency and the capacity since the use of the highest transmission modes reduces the time radio resources are utilized to complete a transmission. In LTE, when an uplink cellular data transmission is paused, DCI reports including the $I_{TBS}$ are not transmitted. In this case, we follow LTE processes in cell-selection or handover, and we monitor the RSRP of the downlink reference signals. The paused LTE cellular data transmission is resumed when the average RSRP (*RSRPavg*) over the last *TcellAvg* seconds is higher than the threshold *RSRPthr*.

## 3. Field trials

Trials have been conducted in the city of Elche (Alicante, Spain) to demonstrate the benefits of opportunistic MCN communications compared to conventional cellular transmissions. The trials used Orange's LTE commercial cellular network under real network and traffic conditions. Trials are conducted in three scenarios for each of the communication modes under evaluation. The selected scenarios are aligned with the scenarios and requirements specified by 3GPP for next generation access technologies [15], in particular with those related to the dense urban scenario. For each trial, the SN has to upload 50MB of UDP data to a server located at the University. A set of over-the-air measurements have been conducted to adjust the parameters of the testbed with the objective to maximize the transmission efficiency. The D2D link was configured following the analysis reported in [13]: *RSSIthr* = -70dBm, *Nb_Rx* = 7, *Nb_BelowThr* = 3, and *T_D2D* = 1s. For the cellular link, $I_{TBS}$*_thr* has been set equal to 18, *RSRPthr* equal to -80dBm and *TcellAvg* equal to 1s. $I_{TBS}$*_thr* has been set equal to 18 following the procedures described in [14] and designed to ensure that active uplink cellular links always use the highest modulation order (and hence transmit at high data rates). *RSRPthr* = -80dBm guarantees that the highest modulation order (and an $I_{TBS}$ equal or higher than 18) is used when a cellular uplink data transmission is started/resumed.

### 3.1. Outdoor trials

The outdoor trials have been conducted at two different locations. At the first one, the SN is initially located behind a building 640m away from the BS under NLOS (Non-Line-Of-Sight) conditions. The SN is initially located 50m away from an intersection where LOS conditions to the BS can be reached. When the tests start, the SN walks towards the intersection and turns around the corner when it reaches the intersection. From this moment onwards, the SN is under LOS conditions with the BS until the test ends. When testing the MCN and opportunistic MCN modes, the RN is initially located 25m away from the SN in the direction of the intersection. At the start of the trials, the RN is under NLOS conditions with the BS and under LOS with the SN. When the trials start, the SN and RN walk towards the intersection and they approximately maintain their separation distance. When the RN reaches the intersection, it turns around the corner and experiences LOS conditions with the BS and NLOS conditions with the SN. When the SN reaches the intersection, it also turns around the corner and experiences LOS conditions with the BS and with the RN. A similar setting is repeated for the second location with a different obstructing building. In this case, the SN is initially located 280m away to the BS under NLOS conditions.

Fig. 2 and Fig. 3 compare the performance achieved with the different communication modes. Fig. 2 represents the average total time needed to upload the 50MB file to the server ('Total' in Fig. 2). This total time represents the time elapsed from the start of the test to the reception of the last packet in the server. The figure also depicts the cellular transmission time ('Cellular' in Fig. 2). This time is computed as the sum of the TTIs (transmission time intervals) during which the SN (for the CC and Opportunistic CC modes) or RN (for the MCN and opportunistic MCN modes) were assigned cellular radio resources to upload data to the BS[2]. The 'D2D' time in Fig. 2 is the time during which the SN transmits data to the RN under the MCN and opportunistic MCN modes. For opportunistic MCN, the 'D2D' time excludes the time during which the D2D transmission was paused. Fig. 3 depicts the cellular spectral efficiency measured in bits/sec/kHz. It is computed as the ratio of the total amount of data uploaded to the cellular BS[3] by the time during which the cellular link is active (i.e. the 'cellular' time in Fig. 2) and by the spectrum (or number of PRBs) used for the uplink cellular transmissions. Fig. 3 also represents the average $I_{TBS}$ experienced during the trials.

Fig. 2 shows that opportunistic CC increases the total transmission time compared to CC. This is the case because the demand-driven opportunistic networking scheme results in that the SN does not start the transmission to the BS until good link quality conditions are experienced.

[2] The difference between the total and cellular times for the SH mode correspond to the sum of the TTIs during which no cellular radio resources were assigned to the SN.

[3] It includes retransmissions, and is hence equal or higher than 50MB.

Such conditions are experienced in this scenario when the SN turns around the corner and it experiences LOS conditions to the BS. However, opportunistic CC reduces by more than 52% the time the SN uses cellular radio resources to upload the file ('Cellular' in Fig. 2) compared to CC. As a result, opportunistic CC increases the spectral efficiency (Fig. 3), and hence the capacity of the cellular network by nearly a factor of 4 compared to CC.

MCN can improve the link budget compared to CC, and therefore reduce the total and cellular transmissions times (Fig. 2), and increase the cellular spectral efficiency (Fig. 3). The D2D link between the SN and the RN reduces the time under which the cellular link experienced NLOS conditions to the BS compared to CC, which favors the use of higher data rates and augmented the average $I_{TBS}$. Additional benefits can be obtained when combining demand-driven opportunistic networking and device-centric wireless communications in the opportunistic MCN mode. Under this mode, the cellular RN-BS link is paused when the RN is under NLOS conditions to the BS. This pause is shorter than with opportunistic CC because the use of an RN reduces the time under NLOS conditions. The D2D SN-RN link is paused when the RN turns around the corner. The opportunistic management of the cellular and D2D links in the opp MCN mode reduces the links' BLER (Block Error Rate), and augments the average $I_{TBS}$ (Fig. 3) and therefore the data rates. This results in that the opportunistic MCN mode achieves the largest reductions in the cellular transmission time (56%) compared to CC[4] while even reducing the total transmission time. Opportunistic MCN also results in the largest increase in the cellular spectral efficiency (a factor of 4.7) compared to CC (Fig. 3), which illustrates the capacity gains that can be achieved by utilizing demand-driven opportunistic networking and device-centric wireless communications. These gains are obtained from the capacity of both technologies to search for the most efficient communication conditions, which includes avoiding NLOS conditions in the selected outdoor scenarios[5].

[4] Opportunistic MCN reduces by 74.5% the cellular transmission time compared to the static configuration of CC. These values are in line with those derived analytically in [12].

[5] The trials under the second scenario (BS is 280m away) have resulted in similar trends to those reported in Fig. 2 and Fig. 3. Smaller gains are observed with respect to CC given that the cellular link quality is higher (even under NLOS) when the BS is closer to the trial area. In this case, opportunistic MCN reduced the cellular transmission time by 47% with respect to CC, and increased the cellular spectral efficiency by a factor of 2.8.

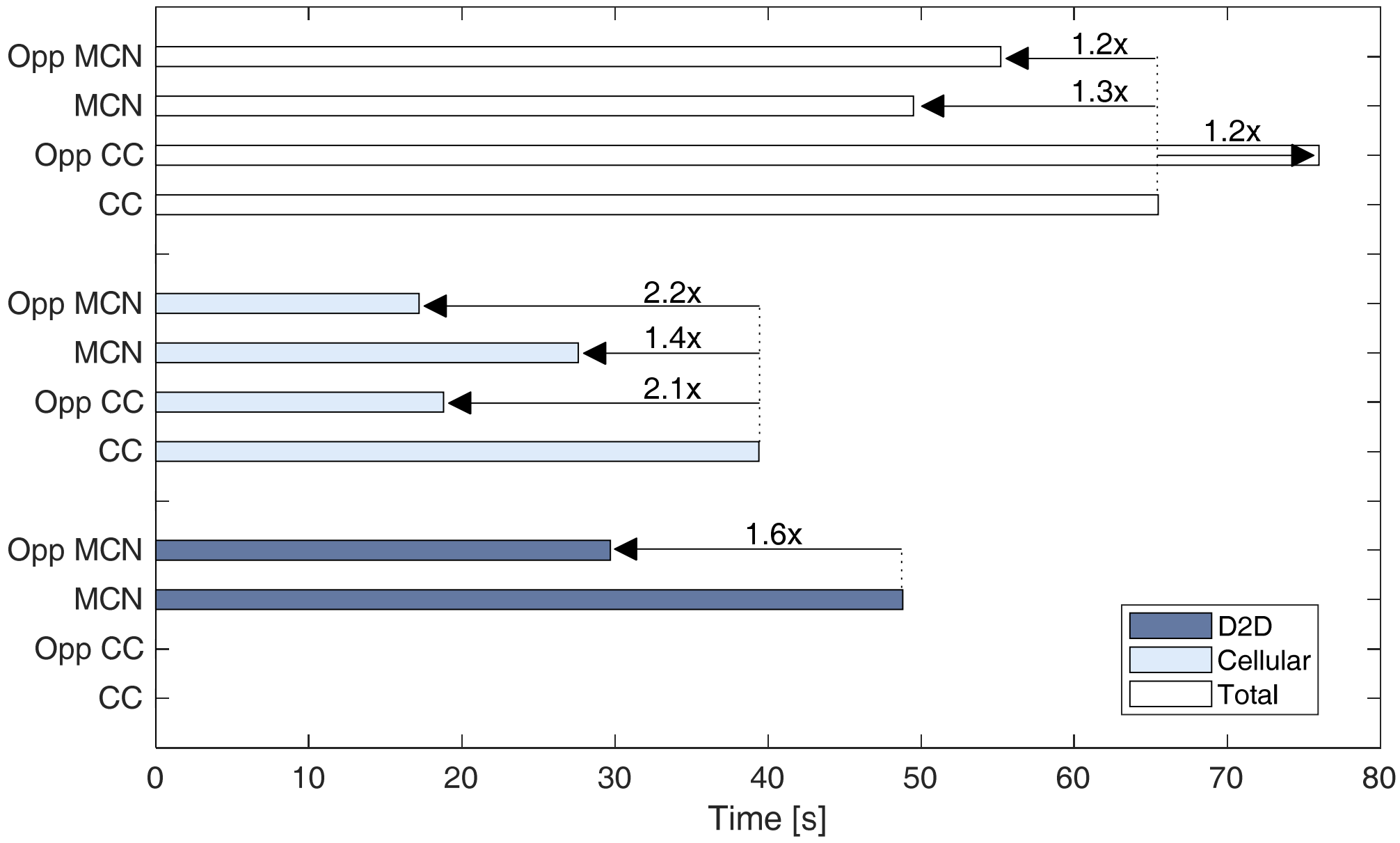

**Fig. 2.** Transmission times for the outdoor trials (BS is 640m away)[6].

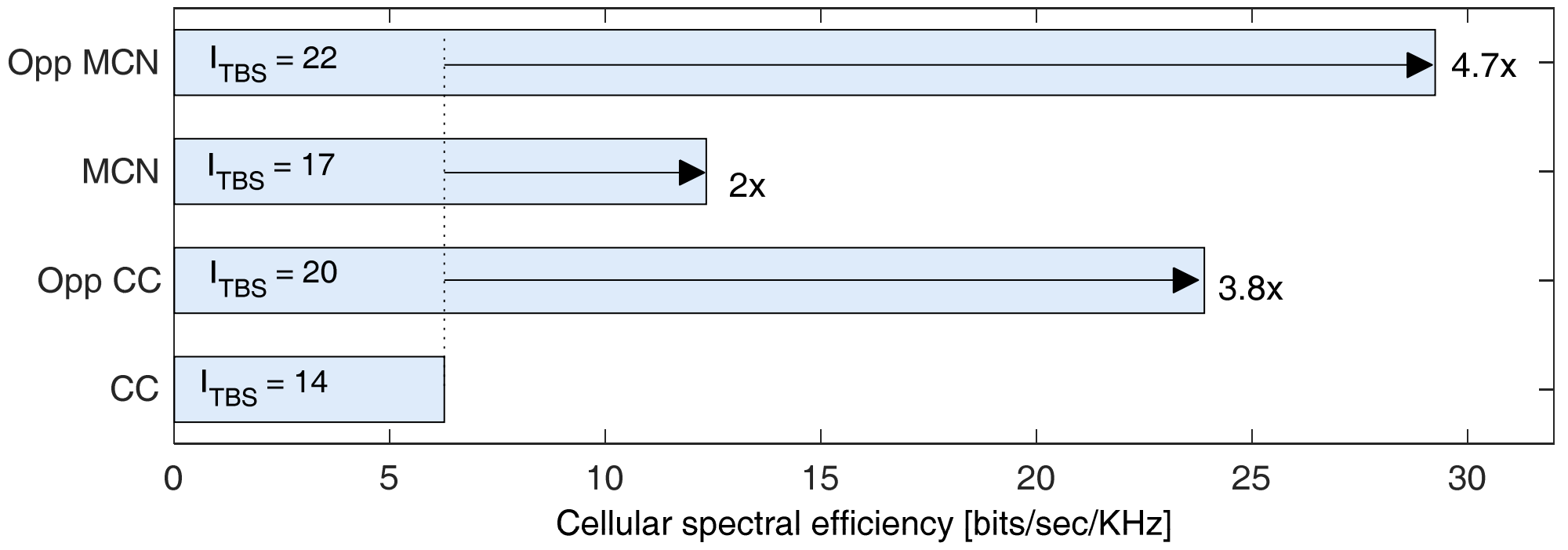

**Fig. 3.** Cellular spectral efficiency (bits/sec/kHz) for the outdoor trials (BS is 640m away)[7].

### 3.2. Indoor-to-outdoor trials

Additional trials have been conducted in indoor-to-outdoor scenarios. Trials have been conducted in three scenarios: two buildings at the University (UMH1 and UMH2) and a shopping center (SC). Users located within these three buildings are served by outdoor BSs that are located 650m, 240m and 500m away, respectively, from the entrance of each building. At the start of the trials, the SN is located inside the building. When the trials start, the SN walks towards the entrance of the building, and the trials finish when the SN has walked out of the building. When trialing the MCN and opportunistic MCN modes, the RN is located 20m away from the SN in the direction towards the entrance of the building. The D2D link between SN and RN experiences LOS conditions. However, pedestrians continuously cross in between the SN and RN. When the trials start, the SN and RN move towards the entrance of the building while maintaining their separation distance. The trials end when both nodes walk out of the building.

---

[6] The worst-case margin of error for the reported average values is below 7.5% with 95% confidence intervals.

[7] The worst-case margin of error for the reported average values is below 7.1% with 95% confidence intervals.

Fig. 4 compares the cellular spectral efficiency of the different communication modes under the three scenarios. Opportunistic MCN results in the highest cellular spectral efficiency in all the scenarios. Opportunistic MCN improves the cellular spectral efficiency of CC by factors ranging from 2.1 to 4.3. The largest gains are observed under the SC scenario that experiences the most challenging conditions for conventional cellular communications with an outdoor BS. During the SC trials, the CC transmissions experienced an average $I_{TBS}$ of 12 compared to 16 and 18 in UMH1 and UMH2 respectively. Opportunistic MCN increases the $I_{TBS}$ to 21 for all the scenarios, and therefore results in the use of higher data rates. The gains achieved by opportunistic MCN come from the combined effect of demand-driven opportunistic networking and device-centric wireless communications. In fact, MCN improves the cellular spectral efficiency compared to CC by a factor ranging between 1.3 and 1.6. These gains are due to the use of the RN that is closer to the entrance, and therefore experiences better cellular link quality levels with the outdoor BS. Opportunistic MCN and opportunistic CC reduce the cellular transmission time (and hence the use of cellular radio resources) compared to CC thanks to the use of higher data rates under more favorable transmission conditions. For example, opportunistic MCN reduces the cellular transmission time of CC by 42% in the SC scenario, and by 35% in the UMH2 scenario. These results are at the origin of the improvement by a factor of 4.3 of the cellular spectral efficiency of opportunistic MCN compared to conventional cellular communications under the most challenging scenario (SC).

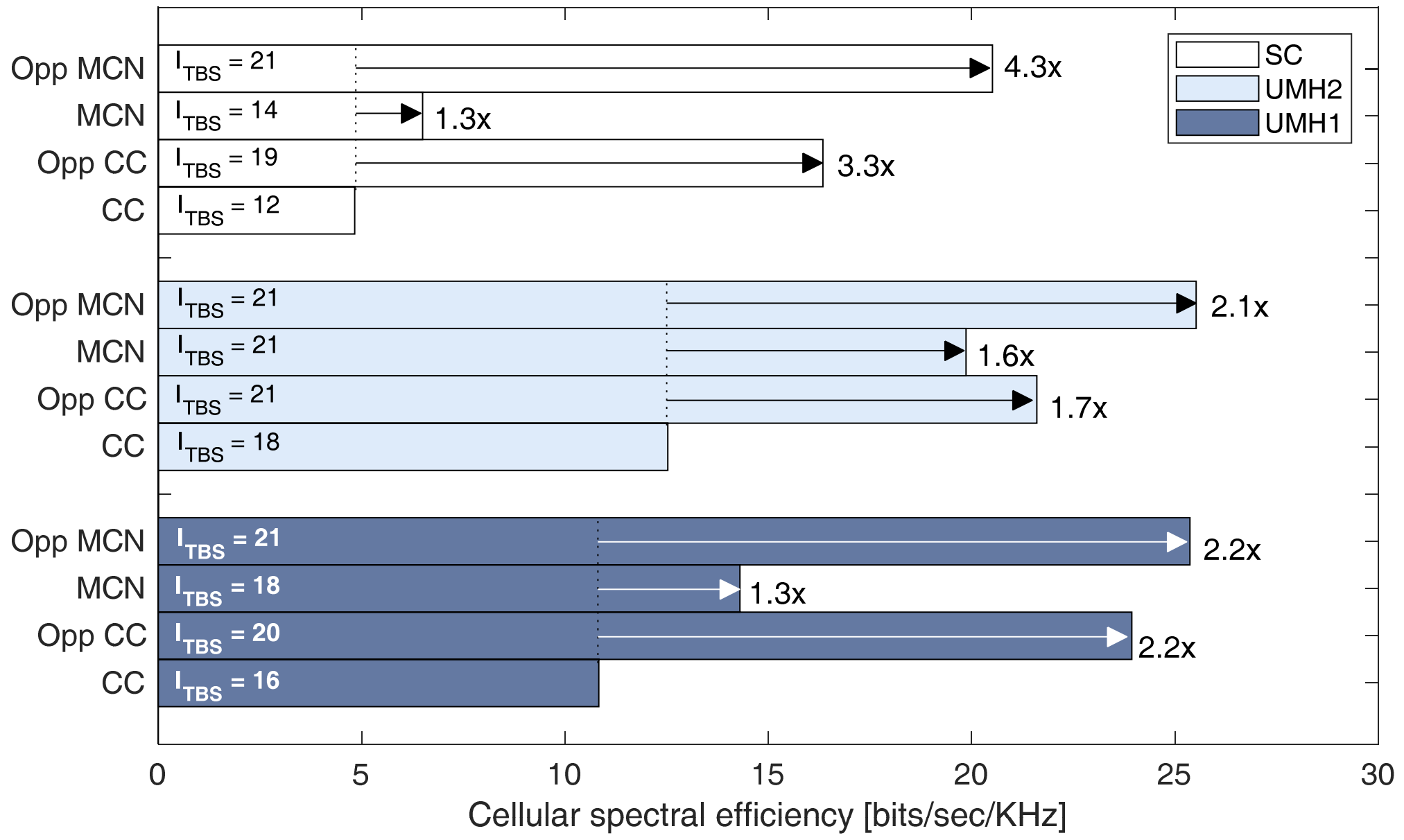

**Fig. 4.** Cellular spectral efficiency (bits/sec/kHz) for the indoor-to-outdoor trials[8].

### 3.3. Vehicles as relays

Vehicles have significant computing capabilities and are becoming increasingly connected. In addition, their mobility offers many possibilities to find the most efficient transmission conditions. In this context, this section explores the benefits of using vehicles as mobile relays in scenarios under LOS and NLOS conditions to the serving BS.

[8] The worst-case margin of error for the reported average values is below 7% with 95% confidence intervals.

*a) LOS conditions*

Trials have been conducted first in a scenario where a pedestrian SN is initially located 440m away from the BS under LOS conditions. When the trial starts, the SN walks through the sidewalk towards the BS. The SN maintains the LOS condition with the BS until the end of the trial. When testing the MCN and opportunistic MCN modes, the RN is mounted in a car. The car is initially located 600m away from the BS and also moves towards the BS. The D2D transmission from the SN to the RN takes place when the car stops in a traffic light close to the location of the SN (approximately 10m). The RN then continues its way and forwards the data traffic to the BS. The cellular link from the RN to the BS also experiences LOS conditions during the tests. Small differences were observed in this scenario between CC and MCN (Table 1) because all nodes are under LOS conditions to the BS, they utilize similar transmission modes ($I_{TBS}$ =14 and $I_{TBS}$ =15, respectively), and the distance travelled by the SN and RN towards the BS during the upload is not significantly different. Higher differences were observed when using opportunistic CC and opportunistic MCN. For these two modes, the SN (pedestrian) or RN (vehicle) do not start the cellular transmission until efficient conditions are experienced (i.e. *RSRPthr* > -80dBm). By doing so, the modes reduce the time they need to utilize cellular radio resources compared to CC (Table 1, by 59.6% and 46.5% respectively) and improve the cellular spectral efficiency (Table 1, by a factor of 4.3 and 3.5 respectively). This is achieved at the expense of a higher total transmission time that is significantly smaller for opportunistic MCN[9] than opportunistic CC. Table 1 shows that opportunistic CC achieved in these trials slightly higher cellular spectral efficiency levels and lower cellular transmission times than opportunistic MCN. This is the case despite the fact that opportunistic MCN transmitted more bits per PRB ($I_{TBS}$ =21) than opportunistic CC ($I_{TBS}$ =20). The cellular transmissions in opportunistic MCN were conducted from a cellular device located inside the car, which augmented the cellular BLER and is at the origin of the differences. Ideally, the cellular transmissions from the vehicle should be carried out using external antennas to avoid the additional propagation losses induced by the vehicle.

*b) NLOS conditions*

In the second scenario, the pedestrian SN is also initially located 440m away from the BS but is under NLOS conditions with the BS. During the trials, the SN walks towards an intersection, turns around the corner when it reaches the intersection, and walks towards the BS. From the intersection onwards, the SN experiences LOS conditions to the BS until the test ends. With MCN and opportunistic MCN, the RN (mounted in a car) is initially located 140m away from the SN (Fig. 5). When the trials start, the RN moves towards the location of the SN and then follows its same trajectory. During the NLOS trials, the RN is mounted on the roof of the car. Fig. 5 depicts the average $I_{TBS}$ indexes that uplink cellular transmissions would experience at the different SN/RN locations during the tests. Fig. 5 shows that CC initially experiences low $I_{TBS}$ (due to the presence of obstructing buildings), but the average $I_{TBS}$ increases when the SN turns around the corner. Opportunistic CC does not start the transmission to the BS until good cellular link quality conditions are experienced, i.e. until the SN reaches the intersection. This results in that opportunistic CC increases the total transmission time compared to CC (Table 1). On the other hand, opportunistic CC reduces the use of cellular radio resources and increases the cellular spectral efficiency significantly (Table 1).

During the MCN tests, the cellular quality between the initial positions of the RN and the SN (Fig. 5) is higher than between the initial position of the SN and the intersection due to the absence of obstructing buildings. This results in that the RN uploads on average 55% of the total file before it reaches the initial location of the SN. When utilizing the opportunistic MCN mode, the D2D link between SN and RN is initially paused since the average RSSI is below -70dBm. When the RN (vehicle) is closer to the SN and the D2D link quality increases, the D2D link activates and the RN-BS cellular transmission starts. By doing so, opportunistic MCN reduces the D2D BLER compared to MCN, and can still upload to the BS 45% of the file before the RN reaches the initial location of the SN. Then, the opportunistic MCN mode pauses the cellular RN-BS transmission until the RN reaches the intersection and experiences LOS conditions to the BS, which increases the average $I_{TBS}$ (Fig. 5). Overall, Table I shows that the opportunistic MCN mode achieves the largest increase in the cellular spectral efficiency (a factor of 12) and the largest reduction in the cellular transmission time (73.7%) compared to CC[10]. These trials experimentally demonstrate the benefit of using vehicles as relays, and their potential to integrate edge networking capabilities in a beyond 5G framework where smart devices become part of the network fabric.

[9] The total transmission time for opportunistic MCN is influenced by the traffic conditions. During several trials, the RN (vehicle) was stopped in the traffic light for some time.

**Table 1.** Performance in vehicular scenarios[11].

| Scenario | Mode | Total tx time [sec] | Cellular time [sec] | Cellular spectral efficiency [bits/sec/kHz] |
|---|---|---|---|---|
| LOS | Opp MCN | 77.7 | 19.1 | 25.3 |
| | MCN | 42.6 | 35.2 | 8.1 |
| | Opp CC | 216.1 | 14.4 | 31.9 |
| | CC | 50.1 | 35.7 | 7.33 |
| NLOS | Opp MCN | 73.8 | 14.5 | 29.3 |
| | MCN | 61.1 | 20.0 | 14.9 |
| | Opp CC | 108.0 | 19.5 | 20.3 |
| | CC | 55.3 | 55.2 | 2.43 |

[10] The comparison of the results obtained under the LOS and NLOS scenarios in Table 1 confirm the impact of using the cellular device mounted on the roof of the car. When the RN device was inside the car (LOS scenario), opportunistic CC achieved slightly higher cellular spectral efficiency levels than opportunistic MCN. When the RN device was mounted on the roof of the car (NLOS scenario), opportunistic MCN achieves higher cellular spectral efficiency levels.

[11] The worst case margin of error for the reported average values is below 5.8% with 95% confidence intervals.

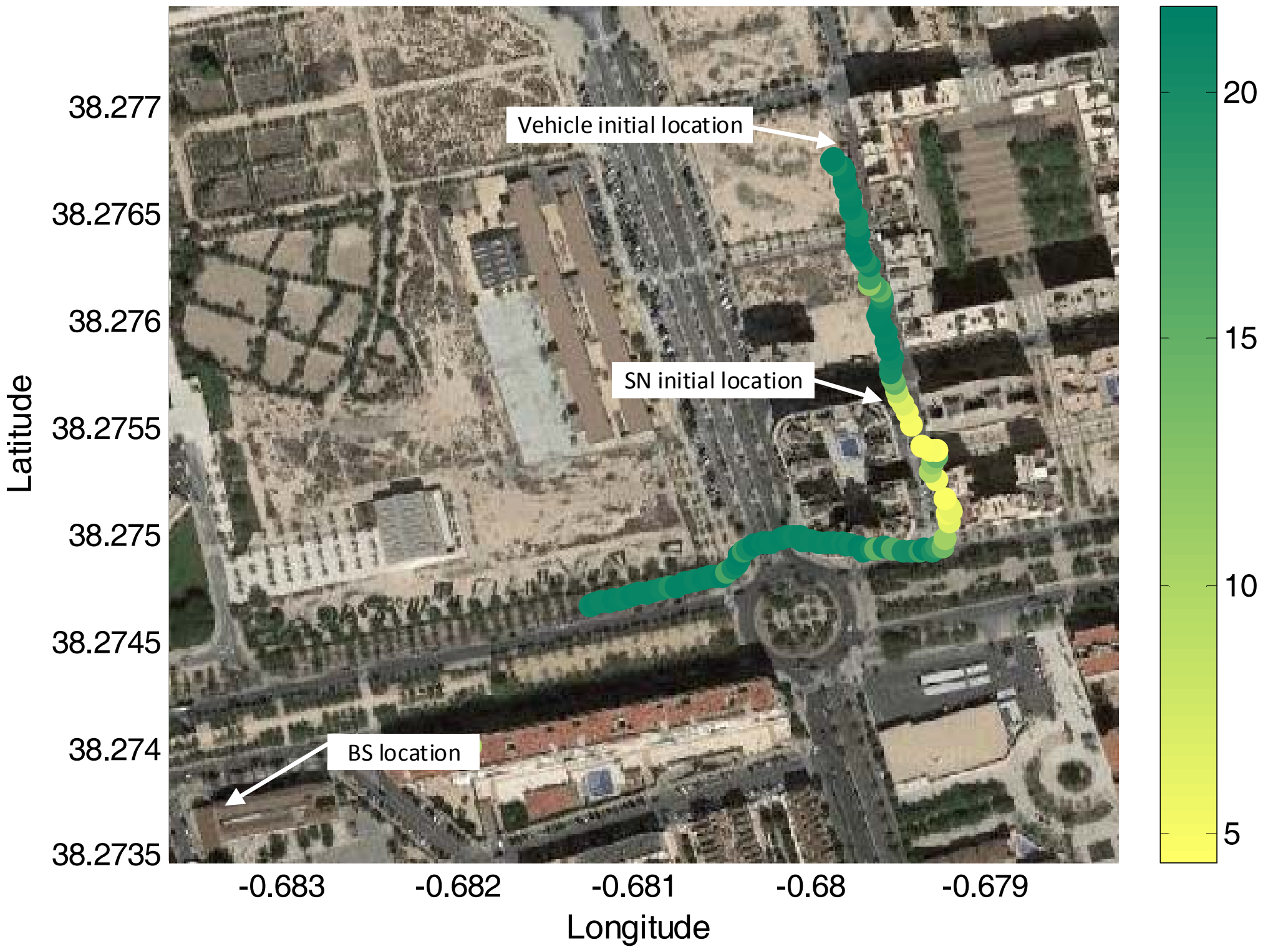

Fig. 5. $I_{TBS}$ under the NLOS vehicular scenario.

## 4. Conclusions

This study has experimentally demonstrated that device-centric wireless communications and demand-driven opportunistic networking can significantly improve the cellular spectral efficiency compared to conventional infrastructure-centric cellular communications. The study has focused on uplink transmissions, even though similar gains and conclusions should be extended for downlink transmissions. The analysis has focused on the uplink in order to have close control of the transmission processes while operating under normal network conditions and without disrupting the normal operation of the base station. The cellular spectral efficiency was improved in the outdoor and indoor-to-outdoor trials by a factor between 4.7 and 2.1 under the evaluated conditions and scenarios. Higher gains were achieved when considering the use of vehicles as relays: the cellular spectral efficiency was improved by a factor of 3.5 and 12 under LOS and NLOS scenarios respectively under the evaluated conditions. These results demonstrate that demand-driven opportunistic networking and device-centric wireless communications can significantly contribute towards achieving the capacity and efficiency goals set up for 5G and beyond. The experiments have been conducted using commercial LTE and 802.11 networks. It is important noting that the gains have been achieved without modifying the LTE and 802.11 radio interfaces, so the potential of demand-driven opportunistic networking and device-centric wireless communications is not constrained to any particular radio interface or generation of (cellular) networks. Both technologies should hence play a fundamental role in the development of 5G and beyond, and help develop a Beyond 5G vision where infrastructure-centric and device-centric networks coexist in a complementary manner. Within this vision, edge networking capabilities can be moved to smart mobile devices that become part of the network fabric. Future networks could hence transform smart mobile devices into cognitive mobile edge network nodes that opportunistically and locally integrate network management functions to ensure that sufficient resources are placed where the demand arises.

**Acknowledgements**
This work has been supported by the Spanish Ministry of Economy, Industry, and Competitiveness, AEI, and FEDER funds (TEC2017-88612-R, TEC2014-57146-R), and the Generalitat Valenciana (APOSTD/2016/049). The authors would also like to acknowledge the support of Orange in Spain.

**B. Coll-Perales** (bcoll@umh.es) received a Telecommunications Engineering degree in 2008 and a Ph.D. in Industrial and Telecommunications Technologies in 2015, both from the Miguel Hernandez University (UMH) of Elche, Spain. He received the Best Student Award in Telecommunications Engineering and the Extraordinary Doctorate Award. As part of his PhD, he spent three months at the Institute of Telecommunications of King's College London (UK) in 2012 working on the design of efficient opportunistic multi-hop cellular networks. In September 2016, he obtained a postdoctoral fellowship from the Valencia regional government that included a one-year postdoctoral visiting period in WINLAB (Rutgers University, NJ, USA) working on millimeter wave for vehicular communications. He is currently a research fellow at UWICORE working on the design of device-centric technologies for future wireless 5G network and connected vehicles. He has been Track Co-Chair in IEEE VTC2018-Fall, Publicity Co-Chair in IEEE VTC2018-Spring, IEEE 5G World Forum, IEEE PIMRC 2016, and IEEE VTC2015-Spring. He serves as Associate Editor of the International Journal of Sensor Networks and the Telecommunication Systems journal.

**Javier Gozalvez** (j.gozalvez@umh.es) received an electronics engineering degree from the Engineering School ENSEIRB (Bordeaux, France), and a PhD in mobile communications from the University of Strathclyde, Glasgow, U.K. Since October 2002, he is with the Universidad Miguel Hernández de Elche (UMH), Spain, where he is currently Full Professor and Director of the UWICORE laboratory. At UWICORE, he leads research activities in the areas of vehicular networks, device-centric wireless networks for 5G and beyond, and industrial wireless networks. He has received several awards at international and national conferences, the best research paper award from the Journal of Network and Computer Applications (Elsevier) in 2014, and the Runner-up prize for the "Juan López de Peñalver" award of the Royal Academy of Engineering in Spain that recognizes the most notable Spanish engineers aged below 40. He is an expert evaluator for the European Commission and research agencies across Europe. He is an elected member to the Board of Governors (2011-2020) of the IEEE Vehicular Technology Society (IEEE VTS), and was the 2016-2017 President of IEEE VTS. He was an IEEE Distinguished Lecturer for the IEEE VTS, and currently serves as Distinguished Speaker. He served as Mobile Radio Senior Editor of the IEEE Vehicular Technology Magazine, and currently serves on the Editorial Board of Computer Networks. He was the General Co-Chair for the IEEE VTC-Spring 2015 conference in Glasgow (UK), ACM VANET 2013, ACM VANET 2012 and 3rd ISWCS 2006. He also was TPC Co-Chair for 2011 IEEE VTC-Fall and 2009 IEEE VTC-Spring. He was the founder and General Co-Chair of the IEEE International Symposium on Wireless Vehicular communications (WiVeC) in its 2007, 2008, and 2010 editions.

**J. L. Maestre** (jmaestre@umh.es) received a B.Sc. in Telecommunications Engineering from the Universidad Miguel Hernández of Elche (UMH) in 2002 and an Electronics Engineering degree from the University of Valencia (UV) in 2006. After finishing his studies, he worked at the Hardware and Software Maintenance Center (PCMSHS) of the Spanish Army from 2006 to 2011. From July 2011, he is a staff member at the Service of Innovation and Technical Support for Education and Research (SIATDI) of UMH, giving support to the UWICORE research laboratory in different research projects such as ENGINE, 5GEAR, FASyS, Coontroler and CGTC.